\documentclass[a4paper,11pt,usenames,dvipsnames]{article}
\usepackage{pos}
\usepackage{caption}
\usepackage{subcaption}
\usepackage{multicol}
\usepackage{braket}
\usepackage{cancel}
\usepackage{multirow}

\title{The hyperon spectrum from lattice QCD}

\author*[a]{Nolan~Miller}
\author[b]{Grant~Bradley}
\author[c]{M.A.~Clark}
\author[d]{Ben~H\"orz}
\author[d,g]{Dean~Howarth}
\author[e]{Malcolm~Lazarow}
\author[f]{Henry~Monge-Camacho}
\author[a]{Amy~Nicholson}
\author[h,i,l]{Enrico~Rinaldi}
\author[d,g]{Pavlos~Vranas}
\author[e,d,f]{Andr\'e~Walker-Loud}

\affiliation[a]{Department of Physics \& Astronomy, University of North Carolina}
\affiliation[b]{Department of Physics, Brown University}
\affiliation[c]{NVIDIA Corporation}
\affiliation[d]{Nuclear Science Division, Lawrence Berkeley National Laboratory}
\affiliation[e]{Department of Physics, University of California, Berkeley}
\affiliation[f]{Department of Physics, University of Costa Rica}
\affiliation[g]{Physics Division, Lawrence Livermore National Laboratory}
\affiliation[h]{Department of Physics, University of Michigan}
\affiliation[i]{iTHEMS, RIKEN, Japan}
\affiliation[l]{Theoretical Quantum Physics Laboratory, RIKEN, Japan}

\emailAdd{nolan@unc.edu}

\abstract{Hyperon decays present a promising alternative for extracting $\vert V_{us} \vert$ from  lattice QCD combined with experimental measurements.  Currently $\vert V_{us} \vert$ is determined from the kaon decay widths and a lattice calculation of the associated form factor. In this proceeding, I will present preliminary work on a lattice determination of the hyperon mass spectrum. I will additionally summarize future goals in which we will calculate the hyperon transition matrix elements, which will provide an alternative means for accessing $\vert V_{us} \vert$. This work is based on a particular formulation of SU(2) chiral perturbation theory for hyperons; determining the extent to which this effective field theory converges is instrumental in understanding the limits of its predictive power, especially since some hyperonic observables are difficult to calculate near the physical pion mass (e.g., hyperon-to-nucleon form factors), and thus
the use of heavier than physical pion masses is likely to yield more precise results when combined with extrapolations to the physical point.}

\FullConference{%
 The 38th International Symposium on Lattice Field Theory, LATTICE2021
  26th-30th July, 2021f
  Zoom/Gather@Massachusetts Institute of Technology
}

\begin{document}
\maketitle
\section{Background: $\vert V_{us} \vert$ from hyperon decays}

Although QCD conserves flavor, the weak interaction does not, a feature of the Standard Model encoded in the Cabibbo-Kobayashi-Maskawa (CKM) matrix. For reference, a global fit of the CKM matrix yields the following entries\footnote{To simplify the notation, some uncertainties have been rounded up.} \cite{Zyla:2020zbs}
\begin{equation}
	\begin{bmatrix}
	\vert V_{ud}\vert  & \vert V_{us}\vert  & \vert V_{ub}\vert  \\
	\vert V_{cd}\vert  & \vert V_{cs}\vert  & \vert V_{cb}\vert  \\
	\vert V_{td}\vert  & \vert V_{ts}\vert  & \vert V_{tb}\vert
	\end{bmatrix} = \begin{bmatrix}
	0.97401(11) & 0.22650(48) & 0.00361(11) \\
	0.22636(48) & 0.97320(11) & 0.04053(83) \\
	0.00854(23) & 0.03978(82) & 0.99917(04)
	\end{bmatrix}
	\, .
\end{equation}
If the mass and weak bases for quarks were identical, this matrix would be diagonal. Instead we see that the weak interaction mixes flavor among generations and flavors, albeit not equally.

The Standard Model predicts this matrix to be unitary, from which one can derive conditions on the rows and columns of this matrix. We will concentrate on the top-row unitarity condition,
\begin{equation}
  \vert V_{ud}\vert ^2 + \vert  V_{us}\vert ^2 +\vert  V_{ub}\vert ^2 = 1 \, .
\end{equation}
Deviations from unity would suggest the presence of physics beyond the Standard Model.

We make the following observations regarding these matrix elements:

\begin{enumerate}
	\item Of the three matrix elements in this relation, $\vert V_{ud}\vert$ is the most precisely known. Here $\vert V_{ud}\vert$ is extracted using superallowed beta decays, in which one calculates a comparative half-life for some nucleus, which can then be averaged with the comparative half-lives from several different nuclei \cite{Towner:2010zz}; the dominant uncertainty comes from the radiative and nuclear structure corrections as predicted by theory \cite{Zyla:2020zbs}.
	\item Historically $\vert V_{us} \vert$ was determined by assuming SU(3) flavor symmetry, which allowed one to estimate the form factors by relating the decays of different baryons in the baryon octet \cite{Towner:2010zz}. However, as SU(3) flavor symmetry is broken by $\sim$15$\%$, this leads to a comparably poor estimate. These days one determines $\vert V_{us} \vert$ instead by using either leptonic ($K_{\ell 2}$) or semi-leptonic ($K_{\ell 3}$) kaon decays in conjunction with a lattice estimate of the associated form factor(s) \cite{Aoki:2021kgd}.
	\item Finally, the last matrix element in this relation, $\vert V_{ub} \vert$, is determined from semi-leptonic $B$ decays; however, it is largely irrelevant for top-row unitarity tests, as its central value is small enough to be eclipsed by the uncertainty of the other two. Thus the top-row unitarity condition is primarily a test between $\vert V_{ud} \vert$ and $\vert V_{us} \vert$.
\end{enumerate}

\begin{figure}
	\centering
	\includegraphics[width=\textwidth]{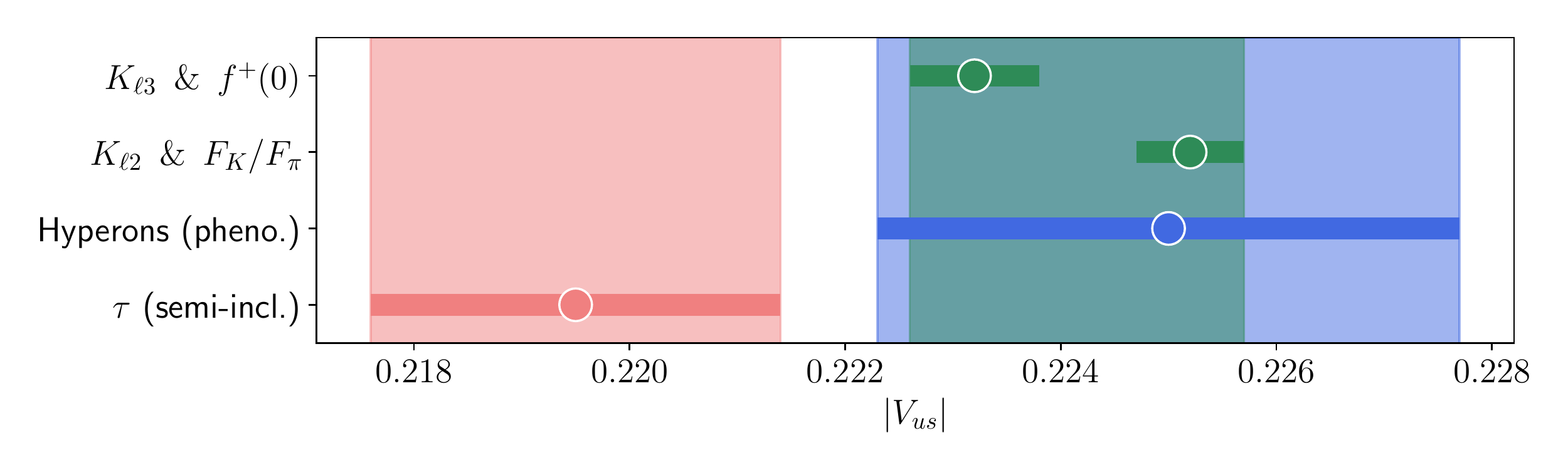}
	\caption{Determinations of $\vert V_{us} \vert$ from different sources.
	The two kaon-derived estimates are taken from FLAG \cite{Aoki:2021kgd}; the phenomenological hyperon-derived value is taken from the Particle Data Group \cite{Zyla:2020zbs} (specifically \cite{Cabibbo:2003ea}); the semi-inclusive $\tau$-derived average is taken from the Heavy Flavor Averaging Group \cite{HFLAV:2019otj}.
	Note that $K_{\ell 2} \enspace \& \enspace F_K/F_\pi$ determines the ratio $\vert V_{us} \vert / \vert V_{ud} \vert$, so here we have also assumed the Particle Data Group average for $\vert V_{ud} \vert$. The green band spans the minimum/maximum values of $\vert V_{us} \vert$ from kaon decays.}
	\label{fig:vus_sources}
  \end{figure}

In Fig.~\ref{fig:vus_sources}, we see that the two kaon-derived values for $\vert V_{us} \vert$ differ by roughly $2 \sigma$. Towards resolving this discrepancy, we would like to calculate the transition matrix elements for the hyperons, thereby providing an orthogonal determination of $\vert V_{us} \vert$. This calculation, however, presents its own set of challenges: the signal is baryonic, not mesonic, and thus inherently noisier; and further, unlike the kaon determinations where only a single form factor (or ratio of form factors) need be determined, here there are multiple form factors in hyperon decays that must be accounted for.

To get an idea of how this works, let us write down the transition matrix element $T$ for the semi-leptonic baryon decay $B_1 \rightarrow B_2 + l^- + \overline \nu_l$ \cite{Towner:2010zz}.
\begin{equation}
	T = \frac{G_\text{F}}{\sqrt{2}} V_{us}
	\Bigg[
	\overbrace{\braket{B_2 | \overline u \gamma_\mu \gamma^5 s | B_1}}^\text{\color{ProcessBlue} axial-vector}
	- \overbrace{\braket{B_2 | \overline u \gamma_\mu s | B_1}}^\text{\color{JungleGreen} vector}
	\Bigg]
	\overline l \gamma^\mu (1 - \gamma^5) \nu_l
	\, .
\end{equation}
The transition matrix element can then be related to the decay widths to extract $\vert V_{us} \vert$, which depends on two hadronic matrix elements. By projecting out the Lorentz structure, we obtain the form factors.
\begin{align}
	{\color{ProcessBlue} \braket{B_2 | \overline u \gamma_\mu \gamma^5 s | B_1}}
	&= g_A(q^2) \gamma_\mu \gamma^5
	+ \underbrace{\cancel{\frac{f_\text{T}(q^2)}{2 M} i \sigma_{\mu \nu} q^\nu \gamma^5}}_\text{\color{RubineRed} G-parity} + \frac{f_\text{P}(q^2)}{2 M} q_\mu \gamma^5
	\\
	{\color{JungleGreen} \braket{B_2 | \overline u \gamma_\mu s | B_1}}
	&= g_V(q^2) \gamma_\mu + \frac{f_\text{M}(q^2)}{2 M} i \sigma_{\mu \nu} q^\nu
	+ \overbrace{\cancel{\frac{f_\text{S}(q^2)}{2 M} q_\mu}}^\text{\color{RubineRed} CVC}
\end{align}

In total there are six form factors, though one can reduce the total to four by invoking the conserved vector current (CVC) hypothesis and appealing to $G$-parity \cite{Weinberg:1958ut}. Given recent measurements of the hyperon decay widths from the LHCb experiment \cite{AlvesJunior:2018ldo}, we believe we can extract a competitive hyperon-derived value of $\vert V_{us}\vert$ so long as we can determine the transition form factors to $\sim$1$\%$.

Before we can calculate all the hyperon transition form factors, however, we will undertake a few more modest goals: first we will calculate the hyperon mass spectrum and then the hyperon axial charges. As our final result for these observables will depend upon an extrapolation based upon $SU(2)$ chiral symmetry breaking for hyperons~\cite{Tiburzi:2008bk,Jiang:2009sf,Jiang:2009fa}, it is prudent to study the convergence pattern of this effective field theory (EFT) and to benchmark our results with the experimental measurements of the hyperon masses.
Prior to having precise lattice QCD results for hyperon quantities, $SU(3)$ baryon chiral perturbation theory ($\chi$PT) was utilized to relate the otherwise numerous low-energy-constants (LECs) describing various processes involving hyperons.  However, $SU(3)$ heavy baryon $\chi$PT does not exhibit a converging expansion~\cite{Walker-Loud:2008rui,PACS-CS:2009cvn,Torok:2009dg} (except possibly for limited observables~\cite{Jenkins:2009wv,Walker-Loud:2011yaf}).
Lattice QCD can be used to determine the more extensive set of LECs that arise in $SU(2)$ (heavy) baryon $\chi$PT for hyperons, thus providing the theory with predictive power.
The benefit of checking the heavy baryon $\chi$PT predictions using the masses are twofold: first, experimental measurements are readily available; second, masses are relatively easy to calculate on the lattice.

The next step will be to calculate the hyperon axial charges.
The leading order LECs that contribute to the hyperon axial charges also describe the pion exchange between hyperons as well as the radiative pion-loop corrections to the hyperon spectrum.
Therefore a precise determination of the axial charges will improve the determination of other observables derived from these Lagrangians.
These same hyperon axial charge LECs will also be important for understanding the interaction of hyperons with nucleons relevant to light hyper-nuclei and possibly for understanding the role of hyperons in neutron stars.

\begin{figure}
	\centering
	\includegraphics[width=0.9\textwidth]{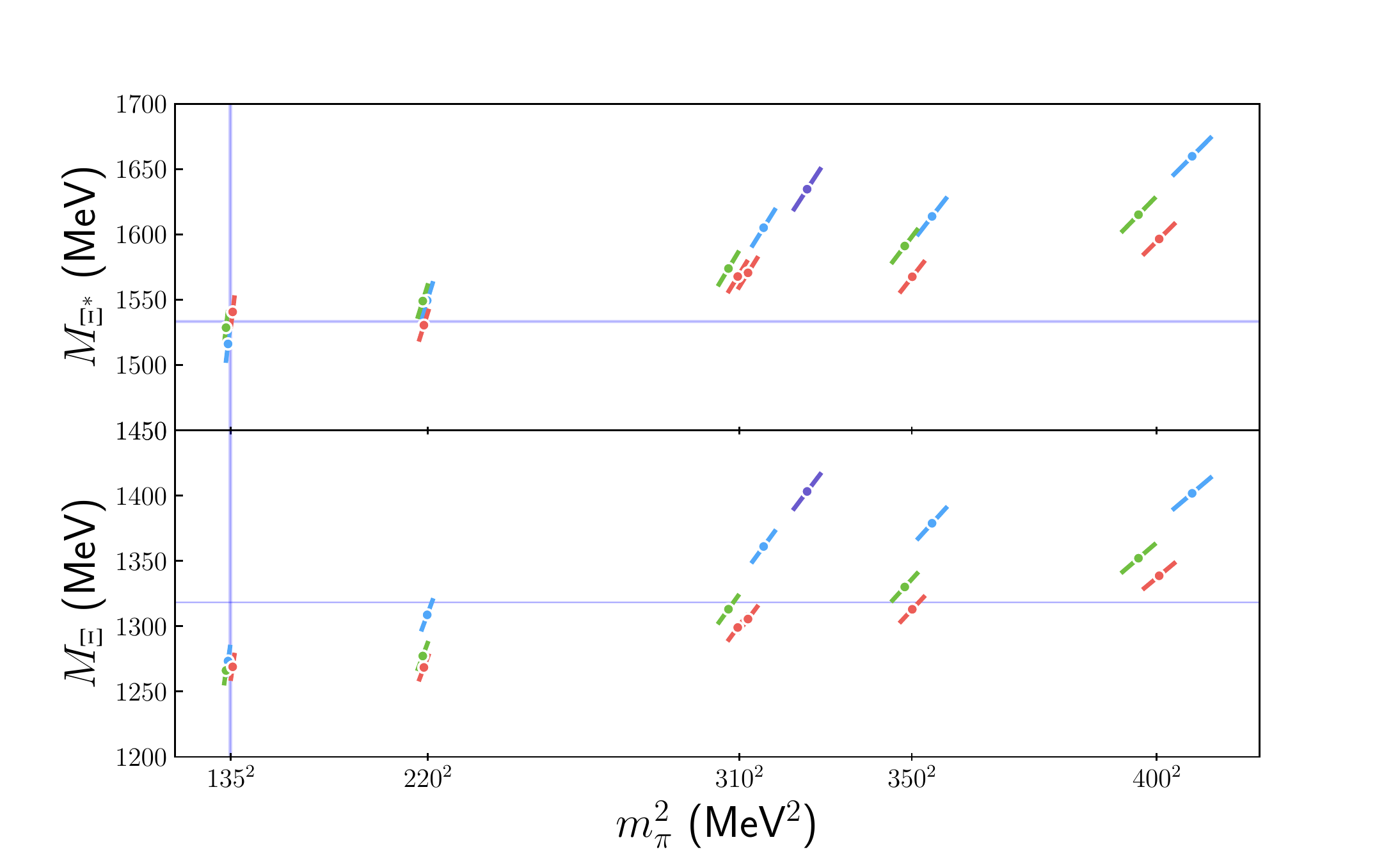}

	\caption{$M_\Xi$ as a function of $m_\pi^2$ for each of our ensembles. Here the lattice spacings range from ${\sim}0.06$ fm (purple) to  ${\sim}0.15$ fm (red). We convert from lattice units to physical units by scale setting with $M_\Omega$ and the gradient flow scale $w_0$ \cite{Miller:2020evg}. The violet bands denote the physical point values of each observable; for $M_\Xi$ in particular, discrepancies between the physical point and the physical pion mass ensembles vanish once corrected for strange quark mistuning and lattice spacing effects.}
	\label{fig:xi_data}
\end{figure}

\section{Project goals \& lattice details}

The eventual goal of this program is to calculate the hyperon transition matrix elements as motivated by the previous section. To perform these calculations, we employ an EFT for hyperons as derived in \cite{Tiburzi:2008bk,Jiang:2009sf,Jiang:2009fa} that relies on heavy baryon $\chi$PT. The first goal of this project is to test the convergence of the EFT employed in this work. To that end, we will first calculate the hyperon mass spetrum, which we determine by taking the chiral mass formula derived from this EFT and extrapolating to the physical point. Later we will calculate the hyperon axial charges and the other transition form factors, which will allow us to determine the transition matrix elements.

The hyperon spectrum has been calculated numerous times, for example in \cite{Durr:2008zz}. There has been comparatively less work on the hyperon axial charges. The first lattice determination of the hyperon axial charges occurred in 2007 but only involved a single lattice spacing \cite{Lin:2007ap}; a calculation involving a physical pion mass ensemble and an extrapolation to the continuum limit didn't occur until 2018 \cite{Savanur:2018jrb}. However, that work only employed a Taylor extrapolation, not a $\chi$PT-motivated extrapolation to the continuum limit. Moreover, our work will benefit from the inclusion of three lattice spacings at the physical pion mass (four in total).

In this work we have used a mixed action with M\"obius domain wall fermions in the valence sector and highly-improved staggered quarks in the sea \cite{Berkowitz:2017opd}, with MILC providing many of the staggered quark configurations \cite{MILC:2010pul,MILC:2012znn}. Fig.\ \ref{fig:xi_data} summarizes the data analyzed so far. Once this project is finished, we will have nearly 30 ensembles with 7 pion masses and multiple volumes.  At the finest lattice spacing, we will have results at pion masses of approximated 220 and 310 MeV.


\section{Extrapolation details}


Let us consider the strangeness $S=2$ hyperons. The chiral expressions for the mass formulae are as follows
\begin{align*}
	M_\Xi^{(\chi)} = &\phantom{+}  M_\Xi^{(0)} + {\color{ProcessBlue} \sigma_\Xi} \Lambda_\chi \epsilon_\pi^2
	& M_{\Xi^*}^{(\chi)} = &\phantom{+}  M_{\Xi^*}^{(0)} + {\color{ProcessBlue} \overline{\sigma}_\Xi} \Lambda_\chi \epsilon_\pi^2
	\\
	& - \frac{3\pi}{2} {\color{JungleGreen} g_{\pi\Xi\Xi}^2} \Lambda_{\chi} \epsilon_\pi^3
	&
	& - \frac{5\pi}{6} {\color{JungleGreen} g_{\pi\Xi^*\Xi^*}^2} \Lambda_{\chi} \epsilon_\pi^3
	\\
	& \qquad- {\color{JungleGreen} g_{\pi\Xi^*\Xi}^2} \Lambda_{\chi} \mathcal{F}(\epsilon_\pi, \epsilon_{\Xi\Xi^*}, \mu)
	&
	& \qquad - \frac{1}{2} {\color{JungleGreen} g_{\pi\Xi^*\Xi}^2} \Lambda_{\chi}  \mathcal{F}(\epsilon_\pi, -\epsilon_{\Xi\Xi^*}, \mu)
	\\
	& + \frac{3}{2} {\color{JungleGreen} g_{\pi\Xi^*\Xi}^2} ({\color{ProcessBlue} \sigma_\Xi} - {\color{ProcessBlue} \overline{\sigma}_\Xi} ) \Lambda_\chi \epsilon_\pi^2
	\mathcal{J} (\epsilon_\pi, \epsilon_{\Xi\Xi^*}, \mu)
	&
	& + \frac{3}{4} {\color{JungleGreen} g_{\pi\Xi^*\Xi}^2} ({\color{ProcessBlue} \overline{\sigma}_\Xi} -{\color{ProcessBlue} \sigma_\Xi} ) \Lambda_\chi \epsilon_\pi^2  \mathcal{J} (\epsilon_\pi, -\epsilon_{\Xi\Xi^*}, \mu)
	\\
	& \qquad + \alpha_\Xi^\text{(4)} \Lambda_{\chi} \epsilon_\pi^4 \log{\epsilon_\pi^2} + \beta_{\Xi}^{(4)} \Lambda_\chi \epsilon_\pi^4
	&
	& \qquad + \alpha_{\Xi^*}^\text{(4)} \Lambda_{\chi} \epsilon_\pi^4 \log{\epsilon_\pi^2} + \beta_{\Xi^*}^{(4)} \Lambda_\chi \epsilon_\pi^4
\end{align*}
where we have defined small parameters
\begin{equation*}
	\epsilon_\pi = \frac{m_\pi}{\Lambda_\chi}
	\qquad
	\epsilon_{\Xi \Xi^*} = \frac{M_{\Xi^*}^{(0)}  - M_{\Xi}^{(0)}}{\Lambda_\chi}
\end{equation*}
with the chiral scale defined as $\Lambda_\chi = 4 \pi F_\pi$.
We have also set the renormalization scale $\mu=\Lambda_\chi$~\cite{Beane:2006kx,Miller:2020xhy}.
The non-analytic $\mathcal{F}$ and $\mathcal{J}$ functions are associated with loop diagrams in the EFT and come at $\mathcal O (m_\pi^3)$ and $\mathcal O (m_\pi^4)$, respectively. Their exact form is not necessary for the high-level discussion in this proceeding. See \cite{Tiburzi:2005na} for details.

From glancing at the chiral expressions, we can immediately glean a few insights. First, in this EFT, baryons of the same strangeness will share many common LECs. Thus we see an immediate advantage of a chiral extrapolation over a Taylor extrapolation: simultaneously fitting both mass formulae will result in more precise determinations of the LECs, which in turn will lead to more precise extrapolations to the physical point. Second, when we later include the axial charges in our analysis, we see that our analysis will benefit twice: once from simultaneously fitting the two and three point functions, thereby improving our determination for the energies on each lattice~\cite{He:2021yvm}; and second when performing the extrapolation to the physical point. Third, as written, the LECs are dimensionless. Indeed, the only dimensionful quantities in the expansion are the constant terms $M_\Xi^{(0)}$ and $M_{\Xi^*}^{(0)}$ and cutoff $\Lambda_\chi$.

\subsection{Results}

\begin{table}
	\small
	\begin{subtable}{0.48\textwidth}
		\begin{align*}
		\begin{array}{rl}
			\\
			+ 1:& \text{Taylor }\mathcal{O}(m^2_\pi) \\
			+ 1:& \text{$\chi$PT } \mathcal{O}(m^3_\pi)\\
			+ 3:& \text{Taylor } \mathcal{O}(m^4_\pi) + \text{$\chi$PT } \left\{ 0,\, \mathcal{O}(m^3_\pi),\, \mathcal{O}(m^4_\pi) \right\} \\
			\hline
			5:& {\color{RubineRed} \text{chiral choices}}
		\end{array}
		\end{align*}
	\end{subtable}
	\hspace{\fill}
	\begin{subtable}{0.48\textwidth}
		\begin{align*}
		\begin{array}{rl}
			\times 5:& {\color{RubineRed}  \text{chiral choices}}  \\
			\times 2:& \left\{ \mathcal{O}(a^2), \mathcal{O}(a^4)\right\}\\
			\times 2:& \text{incl./excl. strange mistuning} \\
			\times 2:& \text{natural priors or empirical priors} \\
			\hline
			40:& \text{total choices}
		\end{array}
		\end{align*}
	\end{subtable}

	\caption{Models employed in this work.}
	\label{tab:models}
\end{table}

\begin{figure}
	\centering
	\includegraphics[width=\textwidth]{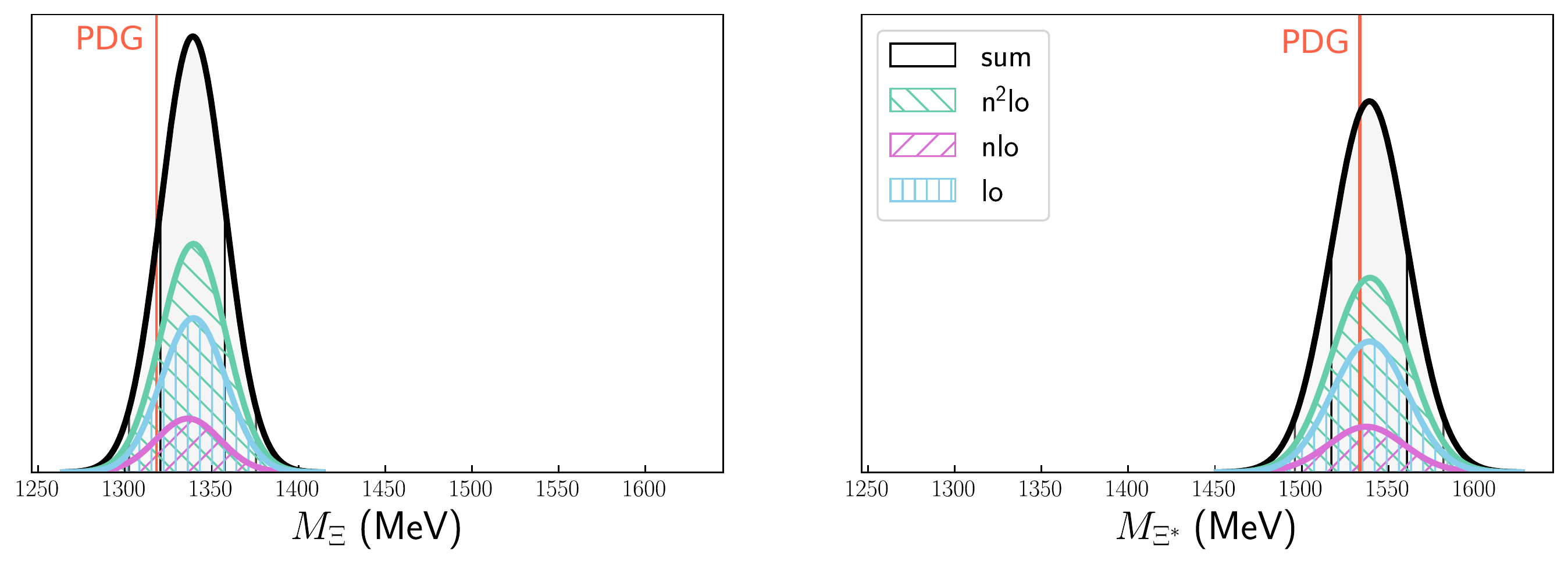}
	\caption{Truncation of the chiral expansion to different orders in the pion mass. We have adopted a ``data-driven'' analysis, i.e.\ we give no \emph{a priori} weight to any of the different chiral models. Although the LO fit only comprises 1/5 of the total models used in this analysis (cf.\ Table \ref{tab:models}), they still contribute more to the model average than the 2/5 of models that truncate at NLO instead. Further, the LO fits contribute almost as much as 2/5 of models that include N$^2$LO terms. The vertical red band is the Particle Data Group average~\cite{Zyla:2020zbs}.}
	\label{fig:histograms}
  \end{figure}

We explore a range of models, summarized in Table \ref{tab:models}, with the models weighted according to their Bayes factors and averaged per the procedure described in \cite{Miller:2020xhy}.
The extrapolation in the pion mass is performed under a range of five choices. We begin by considering a pure Taylor extrapolation to leading order (LO), i.e.\ $\mathcal O (m_\pi^2)$. Next we consider extensions of the LO fit to next-to-leading-order (NLO), i.e.\ $\chi$PT $\mathcal O (m_\pi^3)$ terms. At N$^2$LO, should we choose to include terms of this order, we consider either a pure Taylor term with or without the inclusion of $\chi$PT terms up to $\mathcal O (m_\pi^4)$. Regardless of the pion mass extrapolation, we assume the observables have common LECs per the chiral expression above. Fig.\ \ref{fig:histograms} explores the impact of these different models.

Next we explore corrections specific to the lattice, starting with lattice discretization corrections up to $\mathcal{O}(a^4)$. We also explore the impact of our simulated strange quark mass being slightly mistuned from the physical value.

The priors for the axial charges are set from either experiment or prior lattice calculations \cite{Jiang:2009sf} but with appreciable (20\%) width. The remaining dimensionless LECs are independently priored per the Gaussians $\mathcal N(0, 2^2)$ as is commensurate with "naturalness" expectations. The dimensionful constant terms $M_\Xi^{(0)}$ and $M_{\Xi^*}^{(0)}$ are the exception here and are priored at the physical value of $M_\Xi^{(0)}$ with a 20\% width. We have labeled these the \emph{natural priors}. We have also explored an alternative set of priors derived from the empirical Bayes method.

After model averaging, we report the masses to be
\begin{align}
	M_{\Xi} &= 1339(17)^\text{s}(02)^\chi(05)^a(00)^\text{phys}(01)^\text{M} \text{ MeV} &[1318.28(11) \text{ MeV}]
	\\
	M_{\Xi^*} &= 1542(20)^\text{s}(03)^\chi(06)^a(00)^\text{phys}(03)^\text{M} \text{ MeV} &[1533.40(34) \text{ MeV}]
\end{align}
The values in brackets are the isospin-averaged Particle Data Group values. Here we have separated the errors as induced by statistics (s), chirality ($\chi$), lattice discretization ($a$), physical point input (phys), and model averaging (M). We have not yet calculated the finite volume corrections.

\subsection{The empirical Bayes method}

The empirical Bayes method allows one to estimate the prior distribution from the data; in that sense it is not a truly "Bayesian" approach, as the choice of prior is not data-blind. Nevertheless, it can serve as a useful point of comparison when evaluating the reasonableness of our priors.

Typically when we think of a model for a chiral expression, we imagine this to mean the choice of fit function (e.g., $M = $ ``a Taylor expansion to $\mathcal O(m_\pi^4)$''); however, we can extend the definition of a model to also include the prior. Let us therefore denote $M = \{ \Pi, f\}$ a candidate model for performing the extrapolation of some observable, where $f$ is the extrapolation function and $\Pi$ is the set of priors for the LECs. By Bayes' theorem, the most probable $\Pi$ for a given $f$ and dataset $D$ is

\begin{equation}
	{\color{RubineRed} p(\Pi | D, f)}
	= \frac{{\color{ProcessBlue} p(D |\Pi, f)} {\color{JungleGreen} p(\Pi | f)}}{p(D | f)} \, .
\end{equation}

Here we recognize ${\color{ProcessBlue} p(D |\Pi, f)}$ to be the familiar likelihood function and $p(D | f)$ to be some unimportant normalization constant. The curious term is the hyperprior distribution ${\color{JungleGreen} p(\Pi | f)}$, which parametrizes the distribution of the priors. We restrict our priors to the form $\pi(c_i) = N(0, \sigma^2_j)$ for LEC $c_i$, where the index $j$ denotes some blocking of the LECs. For example, one might use the chiral/discretization split
\begin{equation}
    \Pi = \begin{cases}
        \pi(c_\chi) &= N(0, \sigma^2_1) \\
        \pi(c_\text{disc}) &= N(0, \sigma^2_2)
    \end{cases} \, .
\end{equation}
The hyperprior ${\color{JungleGreen} p(\Pi | f)}$, in this context, parametrizes the $\sigma_j$. We then vary $\sigma_j$ uniformly on the interval $[ \sigma_j^\text{min}, \sigma_j^\text{max} ]$ (in this work, $\sigma_j^\text{min} = 0.01$ and $\sigma_j^\text{max}=100$). As we expect the LECs to be of order 1, we do not expect the optimal values of $\sigma_j$ to lie near the extrema. However, if they do, we should reflect on whether the terms are disfavored by the data ($\sigma_j \sim \sigma_j^\text{min}$) or the LEC is much greater than expected ($\sigma_j \sim \sigma_j^\text{max}$).

Because the hyperprior distribution is uniform, we see that the peak of the posterior ${\color{RubineRed} p(\Pi | D, f)}$  occurs at the peak of the likelihood function ${\color{ProcessBlue} p(D |\Pi, f)}$. Thus the empirical Bayes procedure is straightforward: we find the set of priors that maximizes the likelihood function. But there is one general caveat here. We reiterate that we have blocked the LECs together. One might instead be tempted to optimize each LEC individually; however, this would be an abuse of empirical Bayes---by varying too many parameters, the uniformity assumption can no longer be made in good faith. We emphasize that the empirical Bayes method is not a substitute for careful consideration when setting priors!

\section{Summary \& future goals}

In this work we have calculated the masses of the $\Xi$ and $\Xi^*$ as a first step towards testing the convergence of the hyperon EFT derived in \cite{Tiburzi:2008bk}. The other hyperon masses remain to be calculated. In future work we will use this EFT to calculate the hyperon axial charges and other transition form factors, which will provide an orthogonal method for estimating $\vert V_{us} \vert$.

\section{Acknowledgements}

The work of N.\ Miller was supported in part by a Department of Energy, Office of Science Graduate Student Research award.
A.\ Nicholson was supported by the National Science Foundation CAREER Award Program.
Computing time for this work was provided through the Innovative and Novel Computational Impact on Theory and Experiment (INCITE) program (Summit at OLCF supported by the Office of Science of the U.S. Department of Energy under Contract No.~{DEAC05-00OR22725})
and the LLNL Multi programmatic and Institutional Computing program for Grand Challenge allocations on the LLNL Lassen supercomputer.
QUDA~\cite{Clark:2009wm,Babich:2011np} was used to efficiently solve the MDWF propagators on the GPU-accelerated nodes through the Chroma software suite~\cite{Edwards:2004sx}.
Our calculations were performed with Lalibe~\cite{lalibe}, which links against Chroma, using the \texttt{qedm} branch.

\bibliographystyle{JHEP}
\bibliography{paper}

\end{document}